\documentclass{aa}  

\usepackage{graphicx}
\usepackage{txfonts}
\usepackage{url}
\usepackage{bm}
\usepackage{subfig}

\begin{document}

   \title{Spatially correlated stellar accretion in the Lupus star-forming region}
   \subtitle{Evidence for ongoing infall from the interstellar medium}
   \authorrunning{Winter et al.}
   \author{Andrew J. Winter \inst{1, 2}\fnmsep\thanks{\email{andrew.winter@oca.eu}}
     Myriam Benisty,\inst{1, 2}
     Carlo F. Manara,\inst{3}
     Aashish Gupta\inst{3}
}
           \institute{{Max-Planck Institute for Astronomy (MPIA), Königstuhl 17, 69117 Heidelberg, Germany } \and
           {Universit{\'e} C{\^o}te d'Azur, Observatoire de la C{\^o}te d'Azur, CNRS, Laboratoire Lagrange, 06300 Nice, France}   
           \and {European Southern Observatory, Karl-Schwarzschild-Str. 2, 85748 Garching bei München, Germany}  }

   \date{Received September 15, 1996; accepted March 16, 1997}

% \abstract{}{}{}{}{} 
% 5 {} token are mandatory
 
  \abstract
   {Growing evidence suggests that protoplanetary discs may be influenced by late stage infall from the interstellar medium (ISM). It remains unclear the degree to which infall shapes disc populations at ages $\gtrsim 1$~Myr.}
   {We explored possible spatial correlations between stellar accretion rates in the Lupus star-forming region, which would support the hypothesis that infall can regulate stellar accretion. }
   {We considered both the `clustered' stars towards the centre of Lupus 3, and the `distributed' stars that are more sparsely distributed across the Lupus complex. We took the observed accretion rates in the literature and explore spatial correlations. In particular, we tested whether the clustered stars exhibit a radial gradient in normalised accretion rates, and whether the distributed stars have spatially correlated accretion rates.}
   {We found statistically significant correlations for both the clustered and distributed samples. The clustered sample exhibits higher accretion rates in the central region, consistent with the expected Bondi-Hoyle-Lyttleton accretion rate. Stars that are spatially closer among the distributed population also exhibit more similar accretion rates. These results cannot be explained by the stellar mass distribution for either sample. {Age gradients are disfavoured, though not discounted,} because normalised disc dust masses are not spatially correlated across the region. }
   {Spatially correlated stellar accretion rates within the Lupus star-forming region argue in favour of an environmental influence on stellar accretion, possibly combined with internal processes in the inner disc. {Refined age measurements and} searches for evidence of infalling material are potential ways to further test this finding.}


    \maketitle
%
%________________________________________________________________

\section{Introduction}

\label{sec:introduction}

The timescale for planet formation corresponds to the first few million years of stars' lifetimes, while they still host a `protoplanetary disc' of dust and gas \citep[e.g.][]{Haisch01b}. This timescale is comparable to the typical timescale over which stars form from overdensities in the galactic interstellar medium \citep[ISM, e.g.][]{Jeffreson24}. As a result, planet formation in the protoplanetary disc occurs simultaneously with processes governing star formation, in regions of enhanced stellar and ISM density. Theoretical and observational evidence indicates that several processes, including dynamical encounters \citep{Cuello23} and external photoevaporation \citep{WinterHaworth22} can feedback on the planet formation process. As a result, the external star formation environment may contribute to the diversity of exoplanets. However, how exactly the environment may sculpt planetary systems remains poorly understood.

One process that has been suggested as an important driver of disc evolution is late stage infall of gas from the ISM via the Bondi-Hoyle-Lyttleton (BHL) accretion mechanism, which would explain the observed scaling between accretion rates and stellar mass \citep{Padoan05, Throop08, Klessen10}. {However, a lack of apparent correlation between stellar accretion rates and the environment (based on substantial accretion rates in  Trumpler 37), as well as some theoretical problems in redistributing material to the inner disc, led to the widely held conclusion that BHL accretion is not a primary driver of stellar accretion \citep[][see also discussion in Section 4.1 of \citealt{Winter24}]{Hartmann06}.} Recent years have resulted in a wealth of new observational evidence that motivates a second look at this conclusion. In particular, a recent surge in discoveries of examples of extended structures in molecular line emission \citep{Huang2021} and infrared scattered light around class II discs \citep{Benisty2023, Garufi24} have provoked renewed interest in late stage infall. {Theoretical studies \citep{Dullemond19, Kuffmeier20, Kuffmeier23, Hanawa24} have suggest infall is a viable driver of observed structures such as spirals and streamers \citep{Huang2023, Boccaletti2020}, which are often associated with reflection nebulosity \citep{Gupta23}, accretion outbursts \citep{Seba2020,Hales24,Dong2022} and misaligned inner discs \citep{Ginski2021}. The latter misalignments are apparent among a high fraction of discs \citep{Villenave24}. Ongoing infall may also help explain an apparent mass budget problem for protoplanetary discs compared with the observed exoplanet population \citep{Manara18}, and the non-monotonic evolution of disc dust masses with time \citep{Testi16, Testi22, Cieza19, Cazzoletti19, Williams19}. Infall was also suggested as a possible solution for the presence of old but still accreting discs in star-forming regions \citep{Scicluna2014}, and even a partial explanation for chemical inhomogeneities in globular clusters \citep{Bastian13, Winter23}. From a theoretical perspective, recent simulations have suggested that BHL accretion is a substantial source of mass and angular momentum for discs up to $\sim 1$~Myr after the formation of the host star \citep{Kuffmeier23, Pelkonen24, Padoan24}. Warps instigated by late-stage infall may propagate throughout the disc \citep{Kimmig24, Tanious24}, possibly driving instabilities and short-lived high levels of turbulence in the outer disc \citep{Deng20, Fairburn23}. Shadows resulting from misalignment of the inner disc may also drive spirals and turbulence in the outer disc \citep{Montesinos16, Benisty2017, Su2024, Zhang2024}. Even if inner disc material loses angular momentum by some other mechanism, such as magnetic winds \citep[e.g.][]{Bai13, Tabone22}, infall appears a plausible mechanism to replenish this gas, possibly regulating stellar accretion. This replenishment seems particularly plausible given that the near infrared excess, probing inner disc material, is correlated with outer disc substructures such as spirals and shadows \citep{Garufi18}, suggesting that processes that influence one also influence the other. 

The question of the timescale over which late-stage accretion occurs is of critical importance. If the infall onto the disc is only substantial in the first $< 1$~Myr of disc evolution, then it may be reasonable to assume that this simply sets initial conditions, as conventionally assumed in theoretical studies of planet formation. On the other hand, if substantial infall occurs over the entire disc lifetime, disc replenishment may be a critical ingredient for planet formation models. Observationally, examples of infall have not only been uncovered for very young stars that are embedded in their natal environment \citep{Jorgensen09, Maury19, Pineda20, Codella24}, but also stars older than $ 1$~Myr such as AB Aur \citep{Nakajima95, Grady99,Fukagawa04}, DR Tau \citep{Mesa22}, S CrA \citep{Gupta24} and SU Aur \citep{Ginski2021}. \citet{Garufi24} find evidence of interaction between the star-disc system and the ISM around $16$~percent of their sample of discs in Taurus observed in scattered light. From a theoretical perspective, \citet{Winter24} recently estimated that $\sim 20{-}70$~percent of discs in the age range $1{-}3$~Myr should be mostly composed of recently accreted material, based on the timescale for turbulent fluctuations of the ISM (although this finding may be contingent on stellar feedback). Given the potential critical importance to planet formation models, these studies motivate urgent empirical tests as to whether protoplanetary discs are undergoing significant environmental replenishment.

One approach for testing the importance of replenishment is to correlate observed stellar accretion rates with the external environment. In a medium where the gas relative velocity $\Delta v_\mathrm{gas} \gg c_\mathrm{s}$, the sound speed, the BHL accretion rate $\dot{M}_\mathrm{BHL}$ is proportional to the cross-section carved out by the radius $R_\mathrm{BHL} = 2 G m_*/\Delta v_\mathrm{gas}^2$ within which the gas is captured, where $m_*$ is the stellar mass. Then $\dot{M}_\mathrm{BHL} \propto m_*^2 \rho_\mathrm{gas} /\Delta v_\mathrm{gas}^3$, where $\rho_\mathrm{gas}$ is the local gas density. Thus if BHL accretion is substantially altering disc evolution, then the stellar accretion rate $\dot{M}_\mathrm{acc}$ corrected for the stellar mass dependence -- i.e.  $\dot{M}_\mathrm{acc}/m_*^2$  -- may correlate with local density and anti-correlate with relative velocity of gas. However, finding evidence of such correlations is not necessarily a straight forward task for several reasons. Most importantly, we lack complete (3D) position and velocity information for both the stars and gas. For example, we cannot directly map ISM overdensities along the line-of-sight. The effective noise that this introduces into any statistical signal, alongside numerous other potential complicating factors, is problematic when we consider the modest sample sizes of homogeneously determined accretion rates in most star-forming regions. 

In this work, we investigated whether spatial correlations in stellar accretion are evident in existing observational samples. There remain some approaches for searching for such correlations, even in the absence of very large, homogeneous samples of measured accretion rates (and stellar masses) in individual star-forming regions. One approach is to identify overdensities in tracers for ISM dust or gas that coincide with stellar overdensities. Barring improbable projection effects, such alignment would indicate that the local ISM from which the local stars form has not yet dispersed, and therefore stars and gas share a common physical location, not just on the plane of the sky. A second approach for a more dispersed stellar population is to correlate the spatial proximity of neighbouring stars with relative accretion rates. In a sub-structured star-forming region, stars that are closer together in projected separation are also more likely to be close together in (6D) position-velocity space. If the turbulent scale that dominates the fluctuations in local density and velocity of gas on the scale at which they accrete is greater than the interstellar spacing, then stars that are closer together spatially should also have more similar BHL accretion rates.

In the remainder of this work, we present our investigation into possible correlations between stellar accretion and spatial location in the Lupus star-forming region by adopting these approaches. In Section~\ref{sec:approach} we discuss our approach, including the reasons for choosing Lupus and the datasets we employ. In Section~\ref{sec:results} we report on our findings and assess whether there is evidence of spatially dependent stellar accretion within Lupus. We summarise our findings and steps for the future in Section~\ref{sec:conclusions}. 

\section{Approach and data}

\label{sec:approach}

\subsection{Normalised accretion rate}

As discussed in Section~\ref{sec:introduction}, the BHL accretion rate is dependent both on the stellar mass and the properties of the local ISM. In order to search for correlations between stellar accretion $\dot{M}_\mathrm{acc}$ and the environment it is necessary to isolate the environmental contribution. We therefore define the normalised stellar accretion rate:
\begin{equation}
    \dot{\mathcal{M}} \equiv \dot{M}_\mathrm{acc}  m_*^{-2}.
\end{equation}This factors out the stellar mass dependence of the accretion rate that is expected theoretically (if BHL accretion regulates stellar accretion {and until feedback cuts off the accretion flow -- see Appendix G of \citealt{Winter24}}) and found empirically \citep[e.g.][]{Hartmann16,Manara17,Testi22}. We are interested in determining whether $\dot{\mathcal{M}}$ is randomly distributed or is correlated with position. We therefore require a measurement of both stellar mass and accretion rate for a sufficient sample in a local star-forming region with similar ages.

\subsection{Data}

{In this work, we use the database of homogeneously compiled star and disc properties for the PP7 chapter by \citet{Manara23}. This sample is described in detail in Section 2.5 of that chapter, to which we refer the interested reader. In particular, for this work we required the stellar masses and accretion rates for stars in the Lupus region, which are compiled from the data of \citet{Alcala14} and \citet{Alcala17}. We note that these properties do not have associated individual uncertainties, which are only provided as typical values related to the method used to derive the properties. However, in the context of this work we are interested in inferring correlations, which statistically should only be weakened by the scatter introduced by ignored uncertainties. We limited our sample to stars that have measurements or upper limits for the standardised PP7 accretion rates, stellar mass estimates. This leaves a sample of 74 stars across the whole of Lupus. We further cut this sample by cross-matching with the \citet{Luhman_Lupus20} catalogue of 191 kinematically identified Lupus member candidates. Cross-matching with the PPVII catalogue left a sample of 61. {We further reduced this to 57 stars that were described as `on cloud' by \citet{Luhman_Lupus20}, which are those that are spatially co-located with the Lupus 1--4 clouds.} }

\subsection{Lupus star-forming region}

\begin{figure*}
    \centering
    \includegraphics[width=0.8\textwidth]{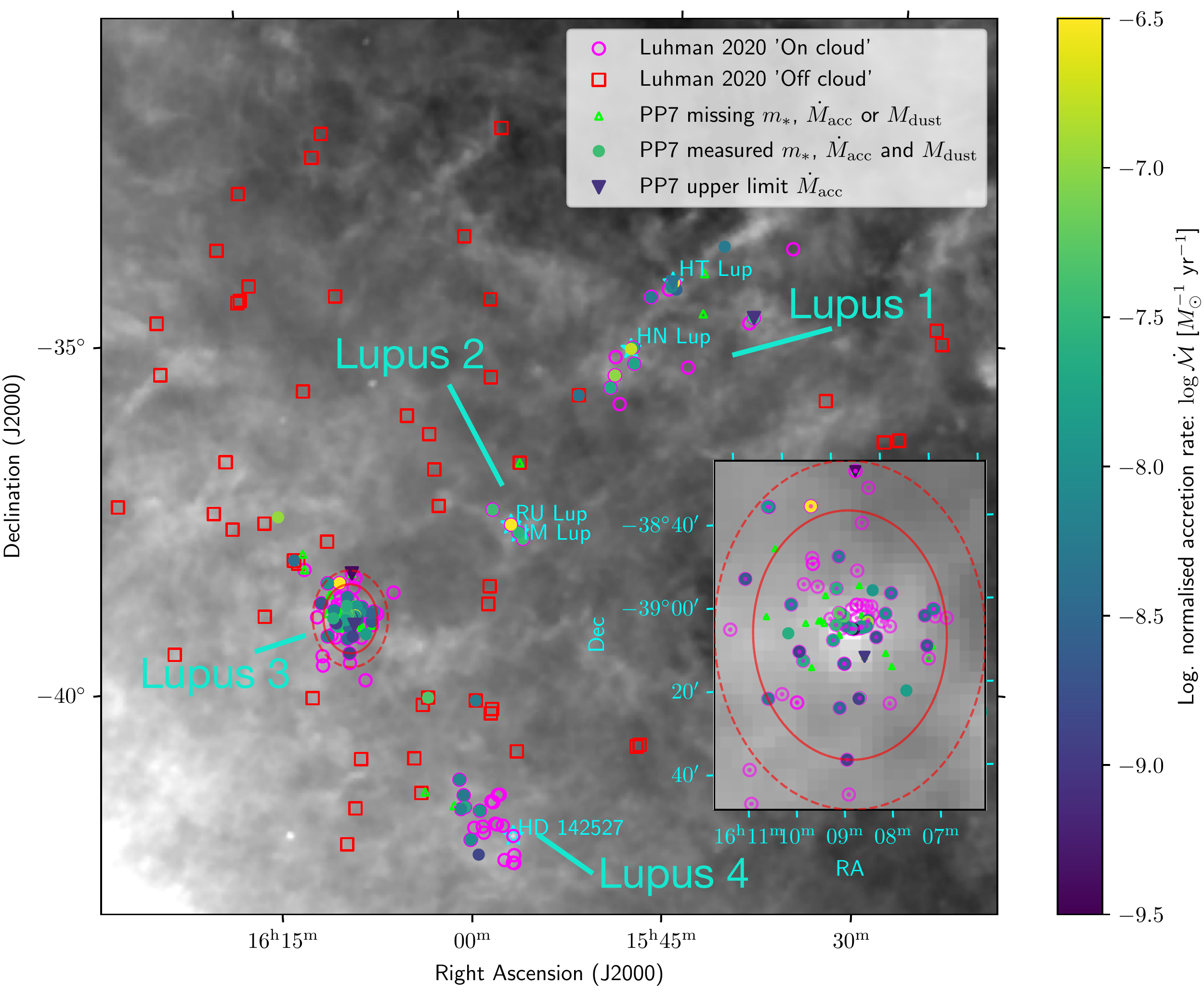}
    \caption{{Distribution of candidate Lupus members with \textit{Gaia} kinematics as identified by \citet{Luhman_Lupus20}, which we show as hollow shapes (fuchsia circles the `on cloud' sample that coincide with Lupus 1-4, and red squares are the `off cloud' sample). The locations of stars with discs included in the PP7 catalogue of \citet{Manara23} but without measurements of stellar mass, accretion rate or disc dust mass are shown as lime triangles. Those stars that have stellar masses, dust masses and accretion rates recorded by \citet{Manara23} are shown as filled coloured circles (or downward triangles for upper limits in accretion rate) that represent the normalised accretion rate according to the colour bar. The greyscale overlay is the IRIS $100$~$\mu$m map of the region \citep{Miville05}. We mark with cyan stars and labels the locations of systems that exhibit some evidence of interaction with the ISM or high levels of outer disc turbulence, discussed in Section~\ref{sec:individ_syst}. We also show a zoom-in on a region centred on Lupus 3, which we define as the `clustered' region (inside the solid red ellipse). Within this circle, we show the `on cloud' sample with fuchsia circles and inner points to make the sample clear in the crowded central region.} }
    \label{fig:lupus}
\end{figure*}

The Lupus star-forming region is one of the closest low-mass star-forming regions, at a distance of $\sim 160$~pc \citep{Lombardi08, Galli_Lupus20}. It is an ideal laboratory to explore whether accretion rates are correlated with local stellar gas, both because its young stellar population is well-studied, and because the discs have intermediate ages \citep[$\sim 1-3$~Myr according to][although see also \citealt{Luhman_Lupus20}, arguing for an older typical age of $\sim 6$~Myr -- we discuss this further in Section~\ref{sec:dmass_test}]{Galli_Lupus20}. If these discs are still substantially influenced by BHL accretion, this would strongly support the importance of this process throughout the disc lifetime. 

Lupus is composed of several molecular clouds, among which the physical conditions of star formation vary. In many of the clouds, star formation is low density and the stellar population is dispersed. However, Lupus 3 is a much denser region, appearing to be a young, low-mass cluster \citep{Schwartz77, Hughes94, Nakajima00}. \citet{Galli_Lupus20} find that the ages of stars in Lupus 3 are typically in the range $2.5-3$~Myr, commensurate with the stars throughout the complex. However, \citet{Rygl13} argue that star formation is also more advanced in Lupus 3 with respect to the other clouds, based on the high star formation efficiency and slowing star formation rate inferred from the comparative dearth of proto- and prestellar sources. 

We show the distribution of stars surrounding the Lupus clouds in Figure~\ref{fig:lupus}. \citet{Luhman_Lupus20} demonstrate that this field contains stars associated with not only the Lupus clouds, but also stars associated with V1062 Sco and Upper Centaurus-Lupus. {However, the PP7 sample already contains disc-hosting stars that have been previously identified as being associated with one of the Lupus clouds. Indeed, the vast majority of the sample are defined as `on cloud' (i.e. spatially coinciding with the Lupus clouds), with only four stars identified as `off cloud' by \citet{Luhman_Lupus20}. In our fiducial sample we retain only the `on cloud' stars, but this does not affect our results.}

For this work, we performed a separate analysis of the clustered star-forming region centred on Lupus 3. We show a zoom-in panel on the central region of Lupus 3 in the bottom right of Figure~\ref{fig:lupus}. The stars within the solid red circle (radius $0.5^\circ$) we label the `clustered' sample in this work. In the clustered sample, there are $32$ stars with stellar mass and accretion rate constraints in the PP7 catalogue with an associated membership inferred from \textit{Gaia} kinematics by \citet{Luhman_Lupus20}. This region is not only densely populated with young stars with respect to the other parts of the cloud complex, but those young stars are centred on a peak in the $100$~$\mu$m IRIS map \citep{Miville05}. This would suggest that the stellar population shares the same spatial location with residual gas from the star formation process. We may therefore expect enhanced BHL accretion in the dense core, with respect to the lower density outer regions --- i.e. a radial gradient in $\dot{\mathcal{M}}$. 

The stars with accretion rate and stellar mass measurements in the less dense parts of the Lupus complex are distributed among the clouds Lupus 1-4. We label these the `distributed' stars, of which 25 kinematically identified `on cloud' Lupus members have measured masses and accretion rates in the PP7 sample. Most of the distributed stars appear to be similar ages $\sim 2{-}3$~Myr \citep{Rygl13}, but none are clearly associated with a stellar overdensity that is coincident with a peak in $100$~$\mu$m emission. Even though we cannot determine the local gas density on a star-by-star basis among the distributed stars, we can assume that the local properties of the turbulent cloud are spatially correlated. In this case, if BHL is contributing to accretion rates, we should expect close neighbouring stars in the distributed population to have more similar $\dot{\mathcal{M}}$. 

{Apart from Lupus, we could also have chosen to study correlations in Taurus or Chameleon I (Cham I), which are also nearby regions that host intermediate age stellar populations. Taurus does not presently contain a sufficient sample of homogeneously derived stellar masses and accretion rates, ruling it out. Cham I does contain such a sample in the PP7 database, however interpreting spatial correlations of accretion rates in this region is not straight forward. Stars in Cham I are distributed along a filament, but there is no clear stellar overdensity (cluster) associated with a monolithic ISM overdensity. Instead, the geometry is complex, and the cloud may be experiencing stellar feedback and dispersal. It is therefore not possible to define `clustered' and `distributed' populations as we have done for Lupus. Such a distinction is necessary since strong local gas density gradients in a bound region may confound efforts to spatially correlate accretion rates. We therefore restricted our attention to the Lupus region.} We focus the remainder of this work on the hypothesised correlations within the `clustered' and `distributed' samples in the Lupus star-forming region.

\section{Results and discussion}\label{sec:results}

\subsection{Clustered stars}
\label{sec:clustered}
\begin{figure*}
     \centering
     \subfloat[][]{ \includegraphics[width=0.5\textwidth]{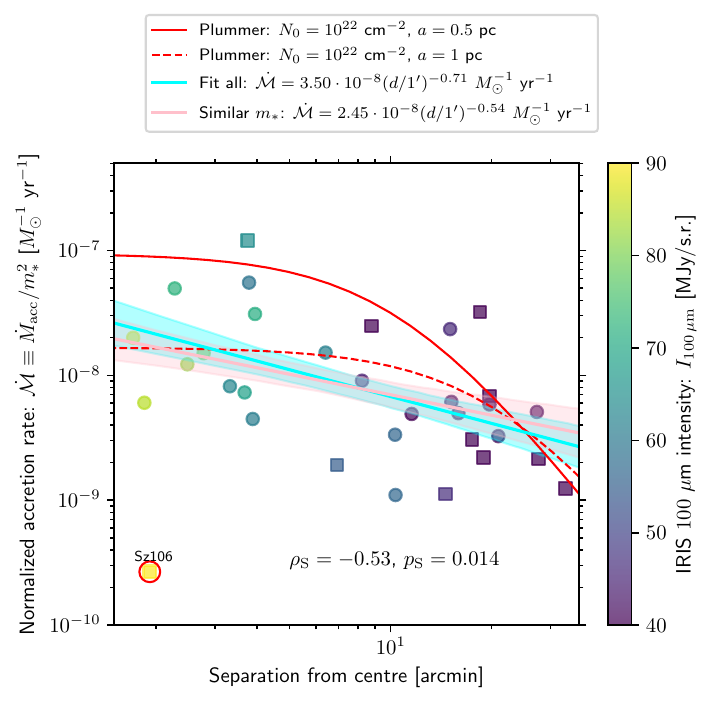}\label{subfig:mdacc_sep}}
     \subfloat[][]{\includegraphics[width=0.5\textwidth]{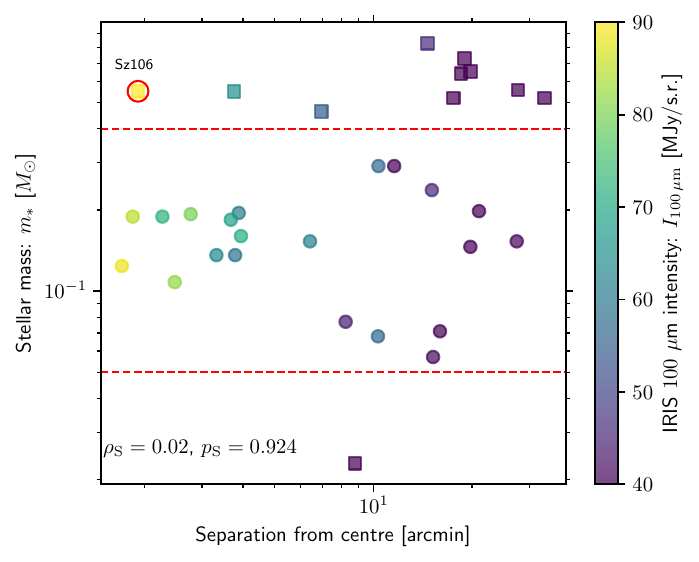}\label{subfig:mstar_sep}}
     \caption{Distribution of normalised accretion rates (Figure~\ref{subfig:mdacc_sep}) and stellar masses (Figure~\ref{subfig:mstar_sep}) as a function of the projected distance from the centre of our clustered sample, which is cut at $0.5^\circ$. Stars used to calculate correlation coefficients are marked as circles, while excluded stars are shown as squares. The latter lie outside the mass limits shown as horizontal dashed red lines in Figure~\ref{subfig:mstar_sep}. All points are coloured by the local $100$~$\mu$m intensity. Figure~\ref{subfig:mdacc_sep} also shows the outcome of fitting a power-law to the full sample (cyan) and the stellar mass restricted sample (pink). We also show the estimated rate of BHL accretion based on two Plummer models (red lines, see Section~\ref{sec:clustered} for details). We highlight with a red circle Sz 106, which is an outlier and sub-luminous star (see Section~\ref{sec:caveats}). }
     \label{fig:mdacc_mstar_sep}
\end{figure*}

We first considered whether there is a radial gradient in $\dot{\mathcal{M}}$ among the clustered population. To refine the nominal centre of our population, we adopted the median RA and Dec from the Lupus sample within our chosen spatial region (red circle, Figure~\ref{fig:lupus}). We then calculated the angular distance from the centre for each of the stars inside this circle. We show the distribution of normalised accretion rates as a function of separation in Figure~\ref{subfig:mdacc_sep}. We exclude from our analysis an outlier, Sz106, which we discuss in Section~\ref{sec:caveats}. We justified our choice to normalise the accretion rate by $m_*^2$ based on the theoretical expectation. However, if this scaling is not the `true' physical scaling, this choice may also introduce spurious correlations if the distribution of stellar masses with measured disc and star properties in the PPVII  sample are not spatially homogeneous. To ensure the robustness of our findings, we constructed a fiducial sample by restricting the stellar mass range $0.05\, M_\odot < m_* < 0.4 \, M_\odot$, as shown in Figure~\ref{subfig:mstar_sep}. These limits are somewhat arbitrary, but motivated by an attempt to maximise the sample while minimising the range in stellar masses. We show in Figure~\ref{subfig:mstar_sep} that stellar masses in this fiducial sample are not spatially correlated.

Among our fidudicial sample, from a Spearman's rank correlation test we find a significant anti-correlation between the separation from the centre and $\dot{\mathcal{M}}$, with significance $p_\mathrm{S} = 0.014$ ($p_\mathrm{S} = 0.023$ if we remove the stellar mass restriction). We also found a similar result by fitting a power-law using \textsc{Linmix}\footnote{\url{https://linmix.readthedocs.io/en/latest/}} \citep{Kelly07}, with a form:
\begin{equation}
    \dot{\mathcal{M}} = A \cdot \left(\frac {d}{1'}\right)^{b},
\end{equation}where $d$ is the angular separation from the centre of the cluster. We find normalisation $A = 3.5 \times 10^{-8} \, M_\odot^{-1}$~yr$^{-1}$ with standard deviation $1.8 \times 10^{-8} \, M_\odot^{-1}$~yr$^{-1}$ and power-law index
$b = -0.56 \pm 0.24$ (pink line in Figure~\ref{subfig:mdacc_sep}). We also find a similar result when including stars of all masses (cyan line in Figure~\ref{subfig:mdacc_sep}) with $b=-0.71\pm 0.23$. 

{We may also ask whether the ISM conditions in Lupus 3 can feasibly drive BHL accretion onto the disc at a rate comparable to the observed stellar accretion rates. While we cannot directly observe the local density and relative gas velocity for individual stars, we can appeal to existing observational constraints to make order of magnitude estimates. The H$_2$ column density in the central regions of Lupus 3 is $N_0 \sim 10^{22}$~cm$^{-2}$, over a region of radius $\sim 1$~pc \citep[see Figure A2 of][]{Rygl13}. Although we do not know the true density profile, from these constraints and by adopting a Plummer density profile we can analytically estimate a reasonable local density and virialised velocity dispersion throughout the cloud. We estimated the normalised BHL accretion rate:
\begin{equation}
\label{eq:MdotBHL}
    \dot{\mathcal{M}}_{\mathrm{BHL}} \approx \pi {G^2 \rho_\mathrm{gas}}/{\sigma_v^3},
\end{equation}where the velocity dispersion $\sigma_v$ and ISM density $\rho_\mathrm{gas}$ are both a function of separation from the centre of the cluster. The result of this exercise is shown by red lines in Figure~\ref{subfig:mdacc_sep}, where we fix $N_0 \sim 10^{22}$~cm$^{-2}$ and adopt $a=0.5$~pc and $a=1$~pc. From this simple estimate, we infer that the expected BHL accretion rate is comparable to or slightly exceeds the typical observed stellar accretion rates. This suggests that, even if material accreted onto the disc does not accrete onto the star with $100$~percent efficiency, environmental infall is a plausible driver of disc evolution and stellar accretion. }

Our findings suggest that stars in the higher density central regions of Lupus 3 accrete at a higher rate than stars in the outer regions. The BHL accretion rate implied by the approximate density and velocity dispersion in the central regions is comparable to the observed accretion rate. Taken together, these findings support the hypothesis that BHL accretion plays a role in regulating stellar accretion, although the substantial scatter may indicate that internal angular momentum transport processes in the inner disc also play an important role. We discuss caveats to this conclusion in Section~\ref{sec:caveats}.

\subsection{Distributed stars}
\begin{figure}
     \centering
     \subfloat[][]{\includegraphics[width=0.8\columnwidth]{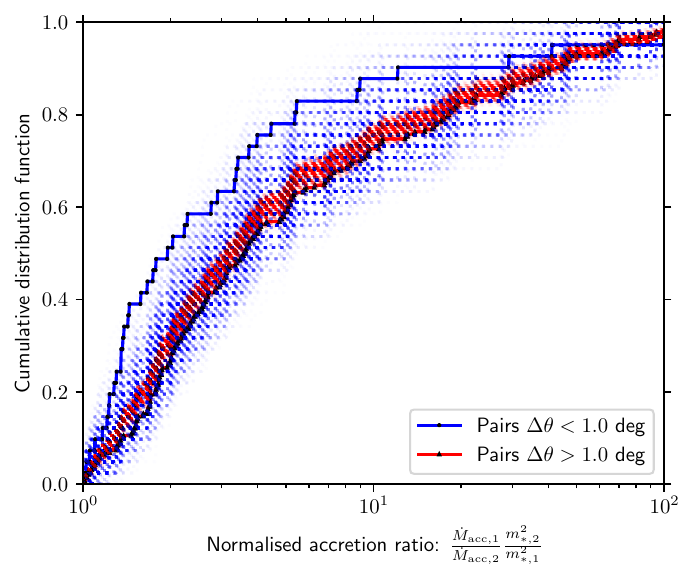}\label{subfig:corr_CDF}}\\
     \subfloat[][]{\includegraphics[width=0.8\columnwidth]{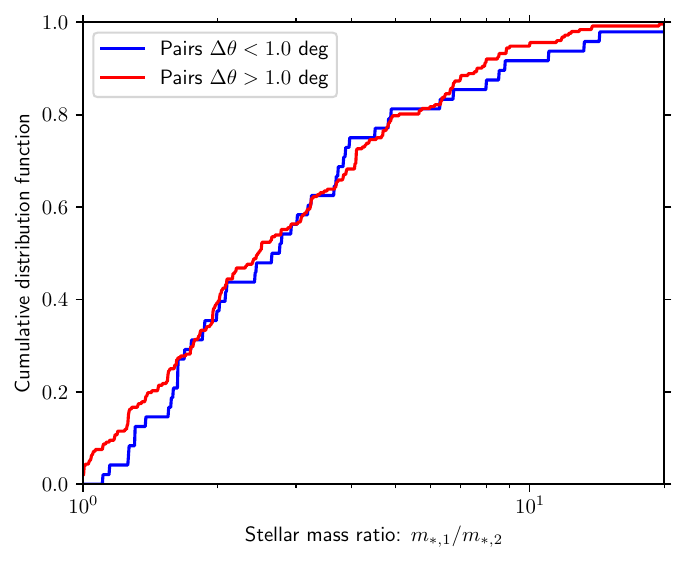}\label{subfig:corr_CDF_mstar}}\\
     \subfloat[][]{\includegraphics[width=0.8\columnwidth]{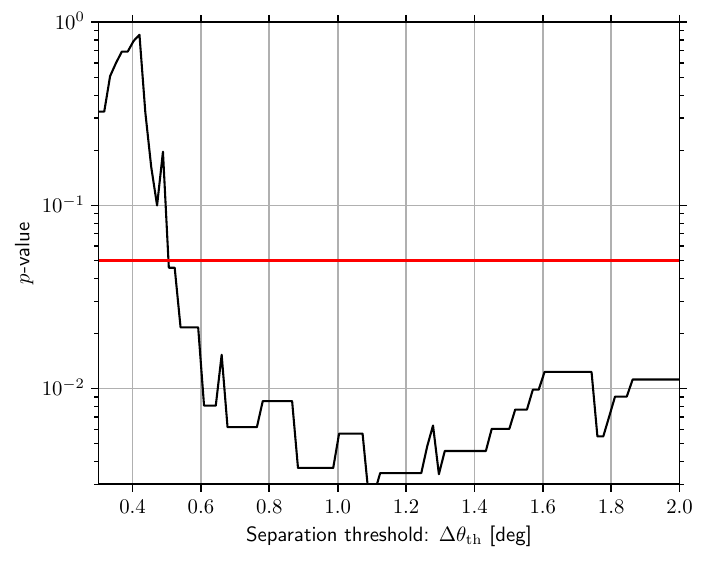}\label{subfig:pDelta}}
     \caption{Outcome of the correlation experiment for the distributed stellar population across the Lupus cloud complex. Figure~\ref{subfig:corr_CDF} shows the cumulative distribution function (CDF) of the ratio of normalised accretion rates between pairs separated by less than (blue) and more than (red) the threshold separation $\Delta \theta_\mathrm{th} = 1^\circ$. Faint blue and red points are bootstrapping experiments where we shuffle the normalised accretion rates between star locations. Figure~\ref{subfig:corr_CDF_mstar} a similar CDF for stellar masses, which are not correlated. Figure~\ref{subfig:pDelta} shows statistical significance as a function of the threshold $\Delta \theta_\mathrm{th}$.    }
     \label{fig:spatial_correlations}
\end{figure}

Outside of the clustered region the stellar population does not have a regular geometry. However, we would still expect stars that are close together to experience more similar BHL accretion rates, since the ISM properties should also be spatially correlated. If BHL accretion influences stellar accretion rates, then such a spatial correlation may also be evident across the measured $\dot{\mathcal{M}}$ distribution. We therefore explored the ratio between the normalised accretion rates between stars $i$ and $j$:
\begin{equation}
    \log \dot{\mathcal{M}}_{\mathrm{r}, ij} = | \log\dot{\mathcal{M}}_{i} - \log\dot{\mathcal{M}}_{j} |,
\end{equation}such that $\dot{\mathcal{M}}_{\mathrm{r}, ij} =\dot{\mathcal{M}}_{\mathrm{r}, ji} >1$ for all $i$, $j$. A large (small) value for $\dot{\mathcal{M}}_{\mathrm{r}, ij}$ means that the normalised accretion rates are very different (similar). We calculated this metric for all the pairs (counting each only once). For this calculation, we exclude upper limits leaving 26 accretion rate measurements, although we confirmed that our conclusions are not affected if we take upper limits as measurements. We then divided up the sample based on whether the pairs are closer than $\Delta \theta_\mathrm{th}=1^\circ$. This choice is arbitrary (we subsequently varied it), but must be larger than typical interstellar spacing while substantially smaller than the scale of the whole region. The outcome of this experiment is shown in Figure~\ref{subfig:corr_CDF}. We find that close pairs have much more similar $\dot{\mathcal{M}}_{\mathrm{r}}$, with significance from a Kologorov-Smirnov (KS) test $p_\mathrm{KS} = 5.7 \times 10^{-3}$. 

Strictly, since each measurement of separation is not independent, the KS test may somewhat overestimate significance. However, we performed a direct bootstrapping experiment by shuffling accretion rates between all individual star locations, measuring the probability of obtaining the observed KS test statistic among these randomly distributed synthetic samples. From this exercise we obtain a similarly significant result $p_\mathrm{boot} = 8.0\times 10^{-3}$, with the bootstrap experiments shown as faint dots in Figure~\ref{subfig:corr_CDF}. We note that the dispersion among the close stars (blue dots) is larger than the distant stars (red dots) because the subsample size is smaller.

We can further ask if there is a bias in terms of the stellar mass distribution within the observational sample. However, we found no significant correlation between stellar mass and position (Figure~\ref{subfig:corr_CDF_mstar}). The (insignificant) differences observed between the stellar mass ratios actually go in the wrong direction, with slightly larger differences in stellar masses among nearest neighbours. 

Finally, we confirmed that our (arbitrary) choice of $\Delta \theta_\mathrm{th}$ is not special. In Figure~\ref{subfig:pDelta} we show KS test results as a function of $\Delta \theta_\mathrm{th}$ obtain a significant result over a wide range of $\Delta \theta_\mathrm{th}$. 

We conclude that the normalised accretion rates among the distributed population are correlated, and that this correlation is not related to an inhomogeneous distribution of stellar masses. We discuss the degree to which this finding supports BHL accretion as a driver of stellar accretion in Section~\ref{sec:caveats}.

\subsection{Age versus environment}
\label{sec:dmass_test}
\subsubsection{Age estimates}
{If angular momentum transport is mediated by internal processes, either viscous or via MHD winds, we expect stellar accretion to decrease with stellar age \citep[e.g.][]{Manara23}. Stellar age gradients have been inferred in other star-forming regions, such as the Orion Nebula cluster \citep[over few parsec scales][]{Hillenbrand97, Getman14,Beccari17}. It is therefore plausible that stars on the outskirts of Lupus 3 are somewhat older than those in the core. Age differences may be even more substantial among the distributed population, which may form from separate gravitationally unstable regions of the ISM at different times. Age gradients generally may therefore partially or fully explain spatial correlations in accretion rates. }

{We estimate what kind of age gradients across Lupus might be needed to explain our results as follows. We first consider Upper Sco, in which stars have typical ages inferred by \citet{Luhman_Lupus20} $\sim 10-12$~Myr \citep[similar to][]{Ratzenbock23}. The typical normalised accretion rate for a star of $\log m_* \sim -0.5 \, M_\odot$ in Upper Sco is $\dot{\mathcal{M}} \approx 1.3 \times 10^{-9} \, M_\odot^{-1}$~yr$^{-1}$, compared to $\dot{\mathcal{M}} \approx 7.9 \times 10^{-9} \, M_\odot^{-1}$~yr$^{-1}$ in Lupus, based on the fits in Figure 7 of that \citet{Almendros-Abad24}. If disc evolution proceeds somewhat self-similarly in this age range, then we would need a substantial age gradient (a factor $\gtrsim 2$) between the inner and outer regions to reproduce the order of magnitude difference in $\dot{\mathcal{M}}$ across our clustered sample (Figure~\ref{subfig:mdacc_sep}). The factor few systematic differences in $\dot{\mathcal{M}}$ across the distributed population (Figure~\ref{subfig:corr_CDF}) are more comparable to the differences between Upper Sco and Lupus, but would still require age gradients comparable to the age differences between the two regions. Overall, it would appear that the systematic variation in stellar age across Lupus would have to be of order the age itself in order to explain the observed correlations without appealing to an environmental origin.}

Stellar ages inferred from isochrone fitting, for example, are notoriously uncertain for individual stars. It is therefore challenging to test whether age gradients may drive our results by directly estimating these ages. {Based on the relative number of stars at different evolutionary stages (defined by the spectral energy distribution -- SED -- classification), \citet{Rygl13} suggested that Lupus 3 is more evolved than Lupus 1, which is in turn older than Lupus 4. However, \citet{Galli_Lupus20} used photometry from \textit{Gaia} DR2, 2MASS, AllWISE and \textit{Spitzer} c2d to fit SEDs to different stellar evolution models, and found that for Lupus 3 and Lupus 4 stellar ages were similar, approximately $2.5 -3.5$~Myr, while sample sizes were prohibitively small across the other clouds. \citet{Luhman_Lupus20} and \citet{Ratzenbock23} used \textit{Gaia} optical photometry to estimate ages averaged over all the clouds, both inferring an older age $\sim 6$~Myr across clouds 1--4. The typical (model-dependent) uncertainty quoted based on the larger sample by \citet{Ratzenbock23} is approximately $\pm 0.5$~Myr, although this is a statistical uncertainty and not a physical spread. The large scale systematic spatial age correlations across the Sco-Cen region hint that typically age differences of several Myr between stellar populations correspond to spatial distances of several tens of parsec.\footnote{\url{https://homepage.univie.ac.at/sebastian.ratzenboeck/wp-content/uploads/2023/05/scocen_age.html}} It would therefore seem unlikely for several Myr age differences across the Lupus clouds (on scales $\sim 10-20$~pc). To yield the correlations we observe, this would also suggest that the intra-cloud age dispersion for the distributed population must be comparatively small, while in Lupus 3 the dispersion is large. These conditions, while not impossible, appear contrived. }

\subsubsection{Disc mass as a proxy}

{We also performed perhaps a more direct test of age differences by considering the properties of our sample. To do so, we reason that if a stellar age gradient were responsible for our findings,} we would expect a spatial correlations of the disc masses. Observationally disc (dust) masses decrease on few Myr timescales \citep[e.g. Figure 5 of][]{Manara23}. We may therefore assume that normalised dust masses are a reasonable proxy for stellar age. Based on this premise, we tested for spatial dust mass correlations across Lupus as follows.

{Analogously to accretion rates, we define the normalised dust mass:}
\begin{equation}
    \mathcal{M}_\mathrm{dust} = M_\mathrm{dust}/m_*^2,
\end{equation}{where $M_\mathrm{dust}$ is the dust mass in the disc as in the PPVII dataset. We have used the same normalisation factor $m_*^2$ in this case, despite there being no clear theoretical motivation for such a scaling with stellar mass. However, observationally the accretion timescale (or $M_\mathrm{dust}/\dot{M}_\mathrm{acc}$) is not strongly dependent on stellar mass \citep{Almendros-Abad24}. }

{To test for correlations among the clustered sample, we restricted the range of stellar masses in our fiducial sample, as for stellar accretion rates. We show the outcome of this exercise in Figure~\ref{fig:mdiscsep}. We found that, regardless of whether we consider the stellar mass-restricted or entire sample, there is no significant correlation between normalised dust mass and distance from the centre of Lupus 3. Fitting power-laws yielded indices $b= -0.14 \pm 0.42$ and $b = -0.25 \pm 0.33$ 
respectively. We conclude that disc masses do not appear correlated with separation from cluster centre.}

\begin{figure}
     \centering
     {\includegraphics[width=\columnwidth]{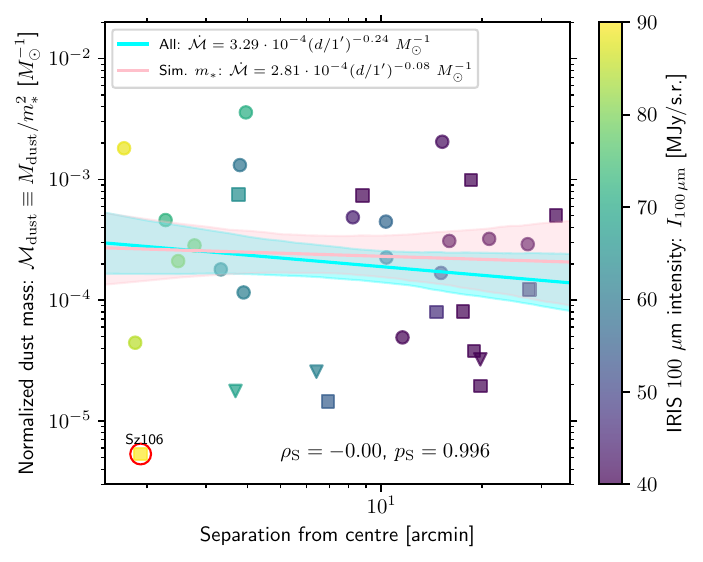}}
     \caption{As in Figure~\ref{subfig:mdacc_sep}, but for normalised disc dust masses rather than accretion rates.}
     \label{fig:mdiscsep}
\end{figure}
\begin{figure}
     \centering
     {\includegraphics[width=\columnwidth]{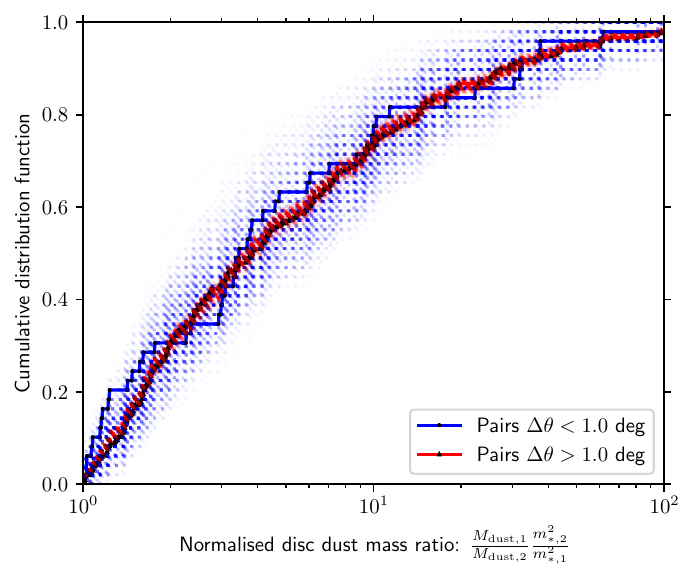}}
     \caption{As in Figure~\ref{subfig:corr_CDF}, but for normalised disc dust masses rather than accretion rates.}
     \label{fig:corr_CDF_Md}
\end{figure}

For the distributed population, as for the clustered population, if age were responsible for the spatial correlations of the normalised accretion rates, we would expect this correlation to be evident also among the normalised disc masses. {Performing a spatial correlation test for the normalised disc dust masses among the distributed population, we find no such evidence (Figure~\ref{fig:corr_CDF_Md}).

We conclude that in Lupus normalised disc (dust) masses do not correlate strongly with spatial location, in contrast to normalised stellar accretion rates. We interpret this as evidence that infall rather than age differences drive the latter correlation. This is because:
\begin{enumerate}
    \item Among the broader sample protoplanetary discs, there is no evidence that accretion rates evolve substantially faster. {In fact, based on Figure 8 of \citet{Almendros-Abad24}, $\mathcal{M}_\mathrm{dust}$ for $\log m_* = - 0.5 \, M_\odot$ decreases from $1.6\times 10^{-2}$ at the age of Lupus to $1.8\times 10^{-3}$ at the age of Upper Sco (a factor $9$ compared to a factor $6$ in $\dot{\mathcal{M}}$). This hints that accretion rates may even evolve marginally slower than dust masses \citep[see also Figure 8 of][]{Manara23}.} Therefore if age gradients were resulting in differences in stellar accretion, they should also result in differences in disc (dust) masses. 
    \item If the origin of our findings is environmental infall, we may not expect such a strong correlation between spatial location and disc mass. This is because disc mass is a time-integrated quantity, which is dependent on the whole disc evolution history. The present-day BHL accretion rate may not be the same as the typical historic rate. By contrast, stellar accretion may be enhanced by turbulence driven by infall onto the outer disc, which is short-lived \citep[e.g.][]{Winter24}. The accretion rate is therefore a more direct probe of the local environment than disc mass.
\end{enumerate}
{Despite the above arguments, we do not categorically rule out age as a factor in our findings. Interpreting our findings as having an environmental origin rests on the assumption that normalised disc dust masses (continuum fluxes) are at least as closely correlated with stellar age as stellar accretion rates are. While present data would suggest this is a safe assumption, large samples with homogeneous age determinations, or correlation experiments with different probes of the total disc mass may help to strengthen or rebuke the infall scenario. {If age is responsible for our findings, this suggests that dust masses must evolve substantially more slowly than accretion rates. In either case, benchmarking models without environmental effects at a specific age to individual star-forming regions may result in incorrect conclusions.}}

\subsection{Sample choice and outliers}
\label{sec:caveats}

\begin{figure}
     \centering
     {\includegraphics[width=\columnwidth]{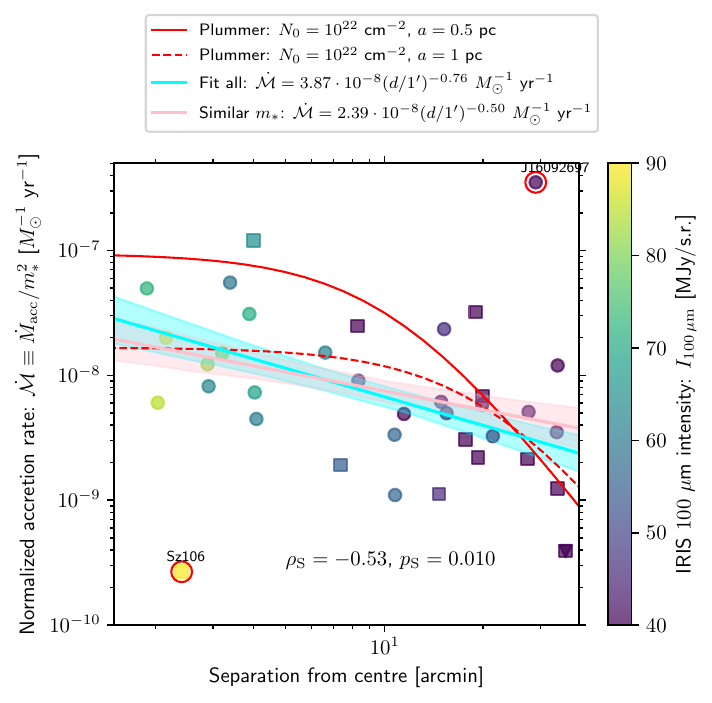}}
     \caption{As in Figure~\ref{subfig:mdacc_sep}, but for a clustered sample contained within a larger radius $0.7^\circ$ from the cluster centre.}
     \label{fig:mstarsep_R07}
\end{figure}

Although our results are statistically significant, the sample size remains small. Our approach has also made it necessary to make some choices, some of which can alter our results. In particular, we have chosen a cut-off radius of $0.5^\circ$ in which to define our clustered sample. This choice is arbitrary, and is made to simply contain the majority of the stars that appear to be approximately spherically distributed around the $100$~$\mu$m peak. 

We therefore tested our conclusions regarding the correlation with separation from the cluster centre against this cut-off radius. First, we find that if we choose instead to limit the sample to those stars within $\lesssim 0.4^\circ$, the significance drops below $2\,\sigma$ due to the reduced sample size. The power-law fit to the between the separation and $\dot{\mathcal{M}}$ remains similar regardless of the cut-off radius down to $\sim 0.2^\circ$. 

If we instead increase the cut-off radius that defines the clustered sample to $0.7^\circ$, as shown by the dashed red circle in Figure~\ref{fig:lupus},
then we include the curious case of J16092697-3836269, with stellar mass $0.15 \, M_\odot$. This is an interesting outlier, which has the largest $\dot{\mathcal{M}}$ across the entire region. It clearly does not fit the trend of the majority of the rest of the clustered sample as shown in Figure~\ref{fig:mstarsep_R07}. 
It also has an estimated dust mass of $M_\mathrm{d} = 1.34 \,M_\oplus$ and an accretion rate of $7.9 \times 10^{-9} \,M_\odot$~yr$^{-1}$. 
Assuming a gas-to-dust ratio of $100$, this implies an accretion timescale of $\tau_\mathrm{acc}\sim 5 \times 10^4$~years, which does not seem compatible with the $\sim 3$~Myr age of the majority of the cluster. Given the anomalously short $\tau_\mathrm{acc}$, we are justified in excluding it from the clustered sample. In this case, we retained a significant correlation for our broader cluster sample (Spearman rank test $p_\mathrm{S} = 0.010$). If we instead included the outlier, this erases the significance according to a Spearman test. This highlights that, while we have justified our choices, we are limited by a relatively small sample size that means a small number of outliers can alter our results. 

{There is a second outlier in the clustered region, which we have excluded in our statistical test due to our choice of mass cut. This is Sz 106, which accretes at a very low level despite being positioned close to the centre of the clustered region. Interestingly, this object has been designated `sub-luminous' by \citet{Comeron03} and \citet{Alcala14}, who pointed out that the star is significantly less luminous than expected for its spectral type. \citet{Alcala14} hypothesise that this is an extinction effect, either due to circumstellar material or an edge on disc. An alternative explanation is an episodic accretion event that could produce a smaller radius, higher temperature, and lower luminosity \citep{Baraffe10}, but this would also imply enhanced surface gravity that is not observed. Whether or not extinction is the origin for the luminosity dip, the accretion rate may be underestimated for this star and we are therefore justified in considering this anomalous. In fact, the known exceptional state of the star in a sense further supports the apparent trend inferred from Figure~\ref{subfig:mstar_sep}, which similarly identifies Sz 106 as an outlier. }

\subsection{Systems with evidence of infall}
\label{sec:individ_syst}

In Figure~\ref{fig:lupus} we have highlighted a number of systems which there is some evidence of infall in the literature. While this list may not be complete, they represent interesting case studies that are plausible evidence of ongoing infall:

\begin{itemize}
    \item {IM Lup} has a very extended ($\sim 400$~au) gas disc with high degree of outer disc turbulence with a turbulent velocity $\sim 0.2{-}0.6\, c_\mathrm{s}$ \citep{PanequeCarreno24, Flaherty24}. IM Lup also has a relatively high normalised accretion rate $\dot{\mathcal{M}} = 2.7\times 10^{-8} \, M_\odot$~yr$^{-1}$. While IM Lup has not previously been suggested as a candidate for infall, it does exhibit an extended and asymmetric halo in $^{12}$CO out to $\sim 1000$~au \citep{Cleeves2016}. This has been previously suggested to be evidence of an externally driven weak photoevaporative wind \citep{Haworth17}, although such asymmetric extended structures could also be interpreted as infalling material. Such infall may also explain the large scale $m=2$ spiral structures observed in the mm-continuum \citep{Huang18}. Modelling is required to examine if infall is a plausible driver of the observed structure. 
    \item {RU Lup} has a disc that exhibits large-scale spiral-like structures, extending out to $\sim 1500$~au, far beyond the $\sim 120$~au Keplerian disc \citep{Huang20}, as well as reflection nebulosity \citep{Gupta23}.  The spiral structures may be signatures of infalling material \citep{Hennebelle2017,Kuffmeier2017}. Interestingly, this large scale is comparable to the extent of the halo around IM Lup. Interpreted in the context of BHL accretion, this would suggest a similar accretion radius, corresponding to similar local gas relative velocity. RU Lup also has a very high $\dot{\mathcal{M}} = 3.3 \times 10^{-7} \, M_\odot^{-1}$~yr$^{-1}$. 
    \item {HD 142527} also exhibits both reflection nebulosity $\sim700$~au spiral structures  \citep{Garg21}, exceeding the disc radius of about 200--300~au \citep{Christiaens2014,Garg21}. HD 142527 also shows a misalignment between the inner and outer disks \citep{Bohn2022}, which may also be a consequence of late infall \citep{Thies2011,Kuffmeier21}. HD 142527 is not included in the PP7 catalogue of star and disc properties, and we therefore did not consider it in the statistics we present in this work. However, based on the accretion rate inferred by \citet{Mendigutia14}, the normalised accretion rate ($\dot{\mathcal{M}} \sim 6 \times 10^{-8} \, M_\odot^{-1}$~yr$^{-1}$) is high. 
    \item {HT Lup} exhibits reflection nebulosity, as well as a crescent-shaped emission extended beyond observed in near-IR polarimetric observations \citep{Garufi2020} and \textit{Herschel} maps \citep{Cieza13}, which may also be a signpost of infall \citep[e.g.,][]{Ginski2021}. However, HT Lup is in fact a triple system \citep[A, B and C; ][]{Correia06}, where the overall luminosity is dominated by HT Lup A \citep{Anthonioz15}, with a mass $1.3\pm 0.2 \, M_\odot$ \citep{Rich22} or $1.32\, M_\odot$ in the PP7 database. Interestingly, HT Lup B hosts a disc that is apparently counter-rotating (or possibly perpendicular) with respect to the disc around HT Lup A \citep{Kurtovic18}. In the PP7 database, HT Lup also only has an upper limit constraint on the stellar accretion rate, such that $\dot{\mathcal{M}} < 4.4 \times 10^{-9} \, M_\odot^{-1}$~yr$^{-1}$. However, recent observations of HT Lup with VLT/MUSE by \citet{Jorquera24} have uncovered evidence of variability in the accretion rate onto HT Lup B (which has a dynamical mass $\sim 0.09 \, M_\odot$) by a factor several, corresponding to $\dot{{M}}_\mathrm{acc} \sim  0.7- 2.9 \times 10^{-9} \, M_\odot$~yr$^{-1}$, or a very large $\dot{\mathcal{M}} \sim 0.9 -3.7  \times10^{-7} M_\odot^{-1}$~yr$^{-1}$. Accretion onto HT Lup A is only somewhat variable ($\sim 30$~percent), with $\dot{\mathcal{M}} \sim 3 \times 10^{-9} M_\odot^{-1}$~yr$^{-1}$. The variable accretion rates in this system may be due to mutual gravitational perturbations between stellar components. However, this would suggest a small physical (line-of-sight) separation that would make explaining counter-rotating discs challenging. HT Lup is also very close to 2MASS J15450887-3417333, which has one of the largest normalised accretion rates ($\dot{\mathcal{M}} = 8.2 \times 10^{-7} \, M_\odot^{-1}$~yr$^{-1}$) across the PP7 sample. 
    \item {HN Lup} exhibits the least compelling evidence of infall due to the absence of necessary ALMA observations. However, it exhibits reflection nebulosity, suggesting that it has substantial local clouds from which it may accrete material \citep{Gupta23}. Like RU Lup, HN Lup also has relatively large $\dot{\mathcal{M}} = 5.1 \times 10^{-8} \, M_\odot^{-1}$~yr$^{-1}$. 
\end{itemize}

To summarise, two of the three cases where we have constraints on $\dot{\mathcal{M}}$ from the PP7 database and evidence of infall have particularly large normalised accretion rates, and one component of the third system (HT Lup B) exhibits a high normalised accretion rate and variability. Added to these three systems, HD 142527 has a high $\dot{\mathcal{M}}$, and IM Lup has a more moderate but still substantial $\dot{\mathcal{M}}$. All of these cases are in the distributed population. This is probably a selection bias. For example, reflection nebulae catalogues used in \citet{Gupta23} are expected to be highly incomplete, particularly in high-density regions (such as Lupus 3) that suffer from higher extinction. A more complete sample of infall candidates is required to statistically correlate the environment with stellar accretion. Nonetheless, the evidence of ongoing infall onto discs around several young stars in Lupus is compelling. Given the short free-fall timescale material at $\sim 1000$~au scales ($\lesssim 10^4 $~yr), this would suggest that this process is either relatively common or sustained by the broader environment even up to the $\sim 1{-}3$~Myr age of Lupus.

\section{Conclusions}
\label{sec:conclusions}

{In this work, we have investigated the possibility that accretion rates in the Lupus star-forming region are spatially correlated, which may be evidence of environmentally regulated stellar accretion. To do so, we divided stars into the clustered population in Lupus 3 centred on a substantial remaining gaseous reservoir, and the distributed population across the rest of the Lupus cloud complex, for which stars are not clearly associated with a particular gaseous overdensity. We found statistically significant correlations both within the clustered population and among the distributed population. In the clustered population, the stellar accretion rates are commensurate with the BHL accretion rate we estimate based on the local gas density. Our results cannot be explained by spatial variations of the stellar mass distribution among the observational sample. Age gradients are also disfavoured, since unlike the normalised accretion rates, the normalised disc masses do not exhibit any similar spatial gradient. However, we cannot categorically rule out the role of age gradients, which may be present particularly among the distributed population.}

{Our results suggest that late stage infall from the ISM onto the protoplanetary disc plays a role in regulating stellar accretion, probably alongside internal processes that drive angular momentum transport in the inner part of the disc. The apparent dependence of star-disc evolution on external environment would not be expected if these latter internal processes were solely responsible for disc evolution. Future studies focused on accurate and homogeneous age determinations or correlating infall occurrence with disc properties may in future strengthen or rebuke this interpretation.}

{Regardless of their origin, our findings are strong evidence that disc populations should not be considered the result of isolated disc evolution from a well-defined initial condition. Even for star-disc systems at the typical age of Lupus ($\sim 1-3$~Myr), stellar accretion rates systematically vary across the region. }

\begin{acknowledgements}We thank the referee for a considered report, which substantially improved the discussion presented in this paper. AJW has received funding from the European Union’s Horizon 2020 research and innovation programme under the Marie Skłodowska-Curie grant agreement No 101104656. This project has received funding from the European Research Council (ERC) under the European Union’s Horizon research and innovation programme for the `PROTOPLANETS' project, grant No. 101002188, and `WANDA', grant No. 101039452. Views and opinions expressed are however those of the author(s) only and do not necessarily reflect those of the European Union or the European Research Council Executive Agency. Neither the European Union nor the granting authority can be held responsible for them. This work has made use of data from the European Space Agency (ESA) mission {\it Gaia} (\url{https://www.cosmos.esa.int/gaia}), processed by the {\it Gaia} Data Processing and Analysis Consortium (DPAC, \url{https://www.cosmos.esa.int/web/gaia/dpac/consortium}). Funding for the DPAC has been provided by national institutions, in particular the institutions participating in the {\it Gaia} Multilateral Agreement.
\end{acknowledgements}

%-------------------------------------------------------------------
\bibliographystyle{aa}
\bibliography{references}

\end{document}